\documentclass[12pt,reqno]{article}
\usepackage{amsfonts}
\usepackage{amsmath}
\usepackage{amsbsy}
\usepackage{graphicx}
\usepackage{amssymb,latexsym}
 \numberwithin{equation}{section}

\begin{document}
 \allowdisplaybreaks[1]
\title{Effective Matter Cosmologies of Massive Gravity I: Non-Physical Fluids}
\author{Nejat Tevfik Y$\i$lmaz\\
Department of Electrical and Electronics Engineering,\\
Ya\c{s}ar University,\\
Sel\c{c}uk Ya\c{s}ar Kamp\"{u}s\"{u}\\
\"{U}niversite Caddesi, No:35-37,\\
A\u{g}a\c{c}l\i Yol, 35100,\\
Bornova, \.{I}zmir, Turkey.\\
\texttt{nejat.yilmaz@yasar.edu.tr}} \maketitle
\begin{abstract}
For the massive gravity, after decoupling from the metric equation
we find a broad class of solutions of the St\"{u}ckelberg sector
by solving the background metric in the presence of a diagonal
physical metric. We then construct the dynamics of the
corresponding FLRW cosmologies which inherit effective matter
contribution through the decoupling solution mechanism of the
scalar sector.
\\ \textbf{Keywords:} Non-linear theories of gravity, massive
gravity, Cosmological solutions
\\
\textbf{PACS:} 04.20.-q, 04.50.Kd, 04.20.Jb.
\end{abstract}

\section{Introduction}
The Fierz-Pauli \cite{fp} massive gravity theory has been extended
to a non-linear Boulware-Deser (BD) \cite{BD1,BD2} ghost-free
level in \cite{dgrt1,dgrt2}. Later this theory is upgraded to
include a general background metric \cite{hr1,hr2,hr3}. The
cosmological solutions of this non-linear and ghost-free theory
has been an active topic of research in recent years
\cite{derham}.

In \cite{constmmgr} we have studied the Einstein solutions of the
so-called minimal sector of the massive gravity. On the other hand
\cite{cosmommgr} was devoted to develop a methodology again for
the minimal theory which would exactly solve the St\"{u}ckelberg
sector by first constructing the solution generating background
(fiducial) metric. The approach in both of these works was to
construct a solution ansatz which would decouple the metric and
the scalar sectors. This enables the determination of the
background metric satisfying the ansatz constraint which leads to
the solution of the scalar sector. In the following work we adopt
a similar formalism for the most general massive gravity theory.
Our objective will be to construct an ansatz which will function
in the same direction. As a solution to this ansatz constraint we
will determine the background metric which will lead us to a class
of solutions of the scalar sector for a given diagonally-formed
physical metric. Such a formulation will replace solving the
scalars from the dynamical field equations by a semi-algebraic
solution of the corresponding ansatz equation. As a physical
consequence of such a solution method admitting decoupling of the
field equations of the metric and the scalar sectors we will be
able to construct the FLRW cosmological dynamics associating the
above-mentioned scalar moduli and the background metric. We will
show that the metric sector thus the cosmological equations get
contributions from an effective matter energy-momentum tensor
which parametrizes the ansatz we consider and which enters into
the metric equation as a remainder of the act of decoupling the
scalars from it. We will also discuss the conservation relation
the effective matter must satisfy. This is a modified version of
the usual energy-momentum conservation. Therefore the effective
ideal fluid appearing in the cosmological equations must be
considered as a non-physical one.

In Section two we construct the necessary ansatz mentioned above
which will decouple the St\"{u}ckelberg scalars from the metric
sector by introducing an effective energy-momentum tensor. In
Section three we will present the explicit solutions of the ansatz
equation for the background metric and the scalars of the theory.
We will also discuss the constraint equation to be satisfied by
the effective matter so that the scalar solutions obtained become
also the solutions of the theory. Finally Section four will be
reserved for the discussion of the dynamics of the corresponding
cosmological solutions of the general massive gravity which
possess effective matter terms as modifications in the Friedmann
and the cosmic acceleration equations.
\section{The ansatz}
 The ghost-free massive gravity action with a general background metric which is coupled to matter can be given as \cite{hr1}
\begin{equation}\label{e1}
 S=-M_p^2\int\bigg[ R\ast 1-2m^2\sum\limits_{n=\texttt{0}}^{\texttt{3}}\beta_{n} e_{n}(\sqrt{\Sigma})\ast 1+\Lambda\ast
 1\bigg]-S_{MATT},
\end{equation}
 where $\beta_n$ are arbitrary coefficients. $M_{p}$ is the Planck mass, $m$ is the graviton mass, $R$ is the Ricci scalar, and $\Lambda$
 is the cosmological constant. Here $\{e_n(\sqrt{\Sigma})\}$ are the elementary
 symmetric polynomials
\begin{subequations}\label{e2}
\begin{align}
e_{0}\equiv e_{0}(\sqrt{\Sigma})&=1,\notag\\
e_{1}\equiv e_{1}(\sqrt{\Sigma})&=tr\sqrt{\Sigma},\notag\\
e_{2}\equiv e_{2}(\sqrt{\Sigma})&=\frac{1}{2}\big((tr\sqrt{\Sigma})^2-tr(\sqrt{\Sigma})^2\big),\notag\\
e_{3}\equiv
e_{3}(\sqrt{\Sigma})&=\frac{1}{6}\big((tr\sqrt{\Sigma})^3-3\,tr\sqrt{\Sigma}\:tr(\sqrt{\Sigma})^2+2\,tr(\sqrt{\Sigma})^3\big),
\tag{\ref{e2}}
\end{align}
\end{subequations}
of the four by four matrix functional $\sqrt{\Sigma}$ in which
\begin{equation}\label{e3}
(\Sigma)^{\mu}_{\:\:\:\nu}=g^{\mu\rho}\partial_{\rho}\phi^a\partial_{\nu}\phi^{b}\bar{f}_{ab}(\phi^{c}),
\end{equation}
with $g^{\mu\nu}$ being the inverse physical metric,
$\{\phi^{a}(x^\mu)\}$ for
$a=\texttt{0},\texttt{1},\texttt{2},\texttt{3}$ the
St\"{u}ckelberg scalars, and $\bar{f}_{ab}(\phi^{c})$ the
background metric. The square root matrix is defined via
$\sqrt{\Sigma}\sqrt{\Sigma}=\Sigma$. The metric equation
corresponding to \eqref{e1} can be derived as \cite{hr1}
\begin{equation}\label{e4}
R_{\mu\nu}-\frac{1}{2}g_{\mu\nu}R-\frac{1}{2}\Lambda
g_{\mu\nu}+\mathcal{T}^S_{\mu\nu}=G_NT^{M}_{\mu\nu},
\end{equation}
where $T^{M}_{\mu\nu}$ is the physical matter energy-momentum
tensor and for later convenience we have written compactly the
contribution of the St\"{u}ckelberg sector as
$\mathcal{T}^S_{\mu\nu}$ which is derived in \cite{hr1} as
\begin{equation}\label{e5}
\mathcal{T}^S_{\mu\nu}=\frac{1}{2}m^2\bigg[\sum\limits_{n=\texttt{0}}^{\texttt{3}}(-1)^n\beta_{n}\bigg(g_{\mu\lambda}Y_{n\,\nu}^{\lambda}(\sqrt{\Sigma})
+g_{\nu\lambda}Y_{n\,\mu}^{\lambda}(\sqrt{\Sigma})\bigg)\bigg] ,
\end{equation}
where
\begin{subequations}\label{e6}
\begin{align}
Y_{0}(\sqrt{\Sigma})&=\mathbf{1}_4,\notag\\
Y_{1}(\sqrt{\Sigma})&=\sqrt{\Sigma}-tr\sqrt{\Sigma}\,\mathbf{1}_4,\notag\\
Y_{2}(\sqrt{\Sigma})&=(\sqrt{\Sigma})^2-tr\sqrt{\Sigma}\sqrt{\Sigma}+\frac{1}{2}\big[(tr\sqrt{\Sigma})^2-tr(\sqrt{\Sigma})^2\big]\mathbf{1}_4,\notag\\
Y_{3}(\sqrt{\Sigma})&=(\sqrt{\Sigma})^3-tr\sqrt{\Sigma}\,(\sqrt{\Sigma})^2+\frac{1}{2}\big[(tr\sqrt{\Sigma})^2-tr(\sqrt{\Sigma})^2\big]\sqrt{\Sigma}\notag\\
&-\frac{1}{6}\big[(tr\sqrt{\Sigma})^3-3\,tr\sqrt{\Sigma}\:tr(\sqrt{\Sigma})^2+2\,tr(\sqrt{\Sigma})^3\big]\mathbf{1}_4,
\tag{\ref{e6}}
\end{align}
\end{subequations}
with $\mathbf{1}_4$ being the four by four unit matrix. Now if we
define the matrix
$[\mathcal{T}^S]^{\mu}_{\:\:\:\nu}\equiv\mathcal{T}^S_{\mu\nu}$
then in matrix notation we can write \eqref{e5} as
\begin{equation}\label{e7}
\mathcal{T}^S=\frac{1}{2}m^2\bigg[\sum\limits_{n=\texttt{0}}^{\texttt{3}}(-1)^n\beta_{n}\big(gY_n
+(gY_n)^T\big)\bigg].
\end{equation}
Now by using the symmetries \cite{bac}
\begin{equation}\label{e8}
g(\sqrt{\Sigma})^n=(g(\sqrt{\Sigma})^n)^T,
\end{equation}
for any integer $n$ and also by referring to the definitions of
the symmetric polynomials \eqref{e2} we can write \eqref{e7} as
\begin{subequations}\label{e9}
\begin{align}
\mathcal{T}^S=m^2\big[&-\beta_1g\sqrt{\Sigma}+\beta_2g(\sqrt{\Sigma})^2-\beta_2\,tr\sqrt{\Sigma}g\sqrt{\Sigma}-\beta_3
g(\sqrt{\Sigma})^3\notag\\
&+\beta_3\,tr\sqrt{\Sigma}g(\sqrt{\Sigma})^2-\frac{1}{2}\beta_3\big((tr\sqrt{\Sigma})^2-tr(\sqrt{\Sigma})^2\big)g\sqrt{\Sigma}\notag\\
&+\big(\sum\limits_{n=\texttt{0}}^{\texttt{3}}\beta_{n}
e_{n}\big)g\big]. \tag{\ref{e9}}
\end{align}
\end{subequations}
To decouple the St\"{u}ckelberg sector from the metric one and
then to solve for the background metric and the scalars of the
theory we now introduce the solution ansatz
\begin{equation}\label{e10}
\mathcal{T}^S=C_1m^2g+C_2m^2\tilde{T}-\frac{1}{2}C_2m^2\tilde{T}^\mu_{\:\:\:\mu}
g+C_2m^2\tilde{\tau},
\end{equation}
where $C_1,C_2$ are arbitrary constants and in matrix sense
$[\tilde{T}]^\mu_{\:\:\:\nu}\equiv\tilde{T}_{\mu\nu}$ is
completely an arbitrary symmetric tensor parametrizing the
solution moduli for which later we will show that it enters into
the metric sector as an effective energy-momentum tensor source.
We have defined its trace as
\begin{equation}\label{e11}
\tilde{T}^\mu_{\:\:\:\mu}=tr(g^{-1}\tilde{T})=g^{\mu\nu}\tilde{T}_{\mu\nu}.
\end{equation}
In \eqref{e10} we have also introduced the matrix
$[\tilde{\tau}]^\mu_{\:\:\:\nu}\equiv\tilde{\tau}_{\mu\nu}$ with
the definition
\begin{equation}\label{e11.1}
\tilde{\tau}_{\mu\nu}=g^{\alpha\beta}\frac{\delta\tilde{T}_{\alpha\beta}}{\delta
g^{\mu\nu}},
\end{equation}
by assuming that the effective energy-momentum tensor will depend
explicitly on the physical metric which will be the case for the
cosmological solutions. Our fundamental motive in proposing this
form of a solution ansatz has a cosmological perspective. The
introduction of the first two terms in \eqref{e10} is for
designing solutions in which the massive sector of the theory
contributes an effective cosmological constant and a non-matter
originated energy-momentum tensor to the metric equations. In this
way the cosmological equations of the theory can be formulated in
the canonical form of the general relativity (GR) ones with
additional effective cosmological constant and matter contribution
which can work as a cure for the dark energy problem of GR
cosmology by admitting self-accelerating solutions. However as we
will discuss next the last two terms in \eqref{e10} are needed for
mathematical consistency so that \eqref{e10} becomes a soluble
ansatz. At this stage there is no guarantee that such an ansatz
may lead to solutions however below we will show that it admits
solutions both in the St\"{u}ckelberg and the metric sectors thus
it is a legitimate one. To generate solutions firstly we have to
make the observation that on the right hand side in \eqref{e9}
there appears
\begin{equation}\label{e12}
 \mathcal{L}_S=\sum\limits_{n=\texttt{0}}^{\texttt{3}}\beta_{n} e_{n},
\end{equation}
which is proportional to the Lagrangian of the St\"{u}ckelberg
coupling of the massive gravity action in \eqref{e1}. Now let us
observe that via \eqref{e1} and \eqref{e4} we have the
inverse-metric variation
\begin{equation}\label{e13}
 \delta\big(2m^2\sqrt{-g}\sum\limits_{n=\texttt{0}}^{\texttt{3}}\beta_{n}
 e_{n}\big)=-\sqrt{-g}\mathcal{T}^S_{\mu\nu}\delta g^{\mu\nu}.
\end{equation}
Upon the introduction of the solution ansatz \eqref{e10} one can
ask the question of what on-shell Lagrangian $\mathcal{L}_{OS}$
this ansatz can be derived from, in other words what the
corresponding Lagrangian level ansatz is. \eqref{e10} and
\eqref{e13} show that $\mathcal{L}_{OS}$ must satisfy the
inverse-metric variation
\begin{subequations}\label{e14}
\begin{align}
 \delta\big(2m^2\sqrt{-g}\mathcal{L}_{OS}\big)=&\sqrt{-g}\big(-C_1m^2g_{\mu\nu}-C_2m^2\tilde{T}_{\mu\nu}-C_2m^2\tilde{\tau}_{\mu\nu}\notag\\
 &+\frac{1}{2}C_2m^2\tilde{T}^\rho_{\:\:\:\rho}
g_{\mu\nu}\big)\delta g^{\mu\nu}\tag{\ref{e14}}.
\end{align}
\end{subequations}
Since for a general Lagrangian the variation solely with respect
to the inverse metric leads to
\begin{equation}\label{e15}
 \delta\big(\sqrt{-g}\mathcal{L}\big)=\sqrt{-g}\big(\frac{\delta\mathcal{L}}{\delta g^{\mu\nu}}-\frac{1}{2}\mathcal{L}g_{\mu\nu}\big)\delta
 g^{\mu\nu},
\end{equation}
we can deduce that
\begin{equation}\label{e16}
 \mathcal{L}_{OS}=C_1-\frac{1}{2}C_2\tilde{T}^\mu_{\:\:\:\mu}.
\end{equation}
We conclude that the solutions which satisfy \eqref{e10} at the
Lagrangian level must lead to $\mathcal{L}_S=\mathcal{L}_{OS}$.
Therefore for the symmetric polynomials of the matrix
$\sqrt{\Sigma}$ we have the on-shell relation
\begin{equation}\label{e17}
 \sum\limits_{n=\texttt{0}}^{\texttt{3}}\beta_{n}
 e_{n}=C_1-\frac{1}{2}C_2\tilde{T}^\mu_{\:\:\:\mu}.
\end{equation}
This analysis shows the necessity of adding the last two terms to
the ansatz \eqref{e10}. If one plans to have the second generic
term in \eqref{e10} then \eqref{e16} is the simplest\footnote{We
did not check any other possible form of Lagrangian which contains
explicitly the effective energy-momentum tensor in it and which
gives such a term. On the other hand we should state that a simple
form for the Lagrangian level ansatz is necessary for deriving
diagonal solutions, and for the simple form of these solutions.}
Lagrangian level ansatz that contains the effective
energy-momentum tensor explicitly in it and produces the second
term in \eqref{e10} upon variation with respect to the inverse
metric. However \eqref{e16} also produces the last two terms of
\eqref{e10}. Hence they must be included in the solution ansatz
when one fixes the on-shell Lagrangian as \eqref{e16} in its
simplest form.
\section{The St\"{u}ckelberg Sector}
In this section we will focus on solving the St\"{u}ckelberg
scalars and the background metric in the solution ansatz
\eqref{e10}. First let us take the trace of \eqref{e10}. If we
multiply both sides by $g$ and then take the trace of the matrix
equation in \eqref{e10} by also using \eqref{e2} after some
algebra we find
\begin{equation}\label{e18}
 4\beta_{0}
 e_{0}+3\beta_{1}e_{1}+2\beta_{2}
 e_{2}+\beta_{3}e_{3}=4C_1-C_2\tilde{T}^\mu_{\:\:\:\mu}+C_2\tilde{\tau}^\mu_{\:\:\:\mu},
\end{equation}
where we introduce the trace
\begin{equation}\label{e18.1}
\tilde{\tau}^\mu_{\:\:\:\mu}=tr(g^{-1}\tilde{\tau})=g^{\mu\nu}\tilde{\tau}_{\mu\nu}.
\end{equation}
Using \eqref{e17} in the above equation will eliminate $e_3$ and
then we can solve for $e_2$. The computation reads
\begin{equation}\label{e18.2}
 e_{2}=\frac{1}{\beta_2}\big[3C_1-3\beta_o-2\beta_1e_1-\frac{1}{2}C_2\tilde{T}^\mu_{\:\:\:\mu}+C_2\tilde{\tau}^\mu_{\:\:\:\mu}\big],
\end{equation}
where we have also used $e_0=1$. Now that we have found $e_2$ in
terms of $e_1=tr\sqrt{\Sigma}$ we can turn our attention to
finding solutions to the ansatz \eqref{e10}. Substituting
\eqref{e9} in \eqref{e10} then multiplying both sides by $g$ and
using \eqref{e17}\footnote{This substitution of the on-shell
Lagrangian in \eqref{e18} and here enables us to eliminate first
$e_3$, then $e_2$ and to obtain the matrix equation \eqref{e19}
which promises simple solutions without the presence of
complicated trace coefficients. Also the simple form of
\eqref{e16} contributes to the simplicity of the $D$- term in
\eqref{e19} which will support our diagonality assumption to
generate solutions in the following analysis.} lead us to the
cubic matrix equation for $\sqrt{\Sigma}$
\begin{equation}\label{e19}
A(\sqrt{\Sigma})^3+B(\sqrt{\Sigma})^2+C(\sqrt{\Sigma})+D=0,
\end{equation}
where we have introduced the four by four matrices
\begin{subequations}\label{e20}
\begin{align}
A&=-\beta_3\mathbf{1}_4,\notag\\
B&=\big(\beta_2+\beta_3e_1\big)\mathbf{1}_4,\notag\\
C&=\big(-\beta_1-\beta_2e_1-\beta_3e_2\big)\mathbf{1}_4,\notag\\
D&=-C_2g^{-1}\tilde{T}-C_2g^{-1}\tilde{\tau}. \tag{\ref{e20}}
\end{align}
\end{subequations}
Although on-shell we have derived $e_2$ in terms of $e_1$ which is
not specified yet we will keep the compact notation of $e_2$ in
the following formulation for the sake of simple appearance of the
equations. To be able to solve \eqref{e19} we will assume that the
physical metric $g$, $\Sigma$ (thus $\sqrt{\Sigma}$), the
effective energy-momentum tensor $\tilde{T}$, and $\tilde{\tau}$
are all diagonal matrices so that \eqref{e19} becomes a diagonal
matrix equation. Our results in the following analysis will
justify that there exist diagonal $\Sigma$ solutions to
\eqref{e19} and we are free to specify the form of $\tilde{T}$ and
we restrict ourselves to the diagonal metric solutions of the
metric sector. This scheme is also conformal with the cosmological
solutions we will consider later. We should state here that the
solution ansatz \eqref{e10} is independent of the diagonality
assumption we will choose for the equation \eqref{e19} in the
following analysis. However for a non-diagonal choice of
$\sqrt{\Sigma}$ equation \eqref{e19} will give a set of coupled
cubic algebraic equations for the entries of the matrix
$\sqrt{\Sigma}$. Also when solved from \eqref{e19} the
non-diagonal form of $\sqrt{\Sigma}$ will lead us to a set of
coupled first-order partial differential equations. In this more
general picture of solutions one can also choose non-diagonal
physical metrics $g$, and the tensors $\tilde{T}$, $\tilde{\tau}$.
Thus more general set of non-diagonal solutions to \eqref{e19} can
be derived by choosing various non-diagonal solution forms of
$\sqrt{\Sigma}$, $g$, and $\tilde{T}$ but in this case one has to
face the difficulty of algebraic and later differential coupling
of equations. On the other hand the diagonality assumption of the
ingredients of \eqref{e19} decouples the algebraic equations for
the entries of $\sqrt{\Sigma}$ first and then the partial
differential equations for the St\"{u}ckelberg fields later. As a
final remark in this direction: assuming diagonality for
$\sqrt{\Sigma}$ but not for $g$ will put extra constraints on the
effective energy-momentum tensor $\tilde{T}$ via \eqref{e19}. Now
let us define
\begin{subequations}\label{e21}
\begin{align}
\Delta_0&=B^2-3AC,\notag\\
\Delta_1&=2B^3-9ABC+27A^2D,\notag\\
\Delta&=-\frac{1}{27}(A^{2})^{-1}\big[\Delta_1^2-4\Delta_0^3].
 \tag{\ref{e21}}
\end{align}
\end{subequations}
If $\Delta>0$ \eqref{e19} has three distinct real roots, if
$\Delta<0$ then it has two complex and one real root, also if
$\Delta=0$ then there are three real roots again with a two-fold
degeneracy. The general solutions of \eqref{e19} can be given as
\begin{equation}\label{e21.1}
\sqrt{\Sigma}=\mathcal{G}_i,\quad\text{for $i=1,2,3$},
\end{equation}
where
\begin{equation}\label{e21.2}
\mathcal{G}_i=-\frac{1}{3}A^{-1}\big[B\mathbf{1}_4+u_i\mathcal{U}+u_i^{-1}\mathcal{U}^{-1}\Delta_0\big],
\end{equation}
with
\begin{equation}\label{e21.3}
u_1=1,\quad u_2=\frac{1}{2}(-1+i\sqrt{3}),\quad
u_3=\frac{1}{2}(-1-i\sqrt{3}),
\end{equation}
and
\begin{equation}\label{e21.4}
\mathcal{U}=\bigg[\frac{\Delta_1+\sqrt{\Delta_1^2-4\Delta_0^3}}{2}\bigg]^{1/3}.
\end{equation}
For the general solutions \eqref{e21.1} of \eqref{e19} the
elementary symmetric polynomial function $e_1$ namely
$tr\sqrt{\Sigma}$ is left undetermined as a solution parametrizing
function. The basic reason for this degree of freedom in the
solution generating scheme we have constructed is the following:
originally there exit two independent parameters namely $e_2$, and
$e_1$ in the solution ansatz \eqref{e10} as we can eliminate $e_3$
by the introduction of the on-shell Lagrangian \eqref{e17}, when
one takes the trace of \eqref{e10} which is equivalent to
\eqref{e19} one obtains \eqref{e18.2} which enables us to express
$e_2$ in terms of $e_1$, on the other hand one may expect to
determine $e_1$ from the explicit solution \eqref{e21.1} by taking
the trace of it as well, however \eqref{e21.1} is algebraically
another way of writing \eqref{e19} which it satisfies as a
solution thus taking its trace would lead us to no additional
identity than \eqref{e18.2}. Therefore we can conclude that in the
solution \eqref{e21.1} $e_1$ remains as a free function to
parametrize the solutions. Below we will see that demanding the
reality of the solutions may bring restrictions on $e_1$ or fix
and determine its functional form in terms of the other free
function parameters of the solutions (i.e. the effective
energy-momentum tensor). Now by assuming that we focus on the real
solutions of \eqref{e19} squaring both sides of \eqref{e21.1}
yields
\begin{equation}\label{e21.41}
\Sigma=g^{-1}f=\mathcal{G}^2,
\end{equation}
where we introduce the pull-back of the background metric
$\bar{f}$ by using the St\"{u}ckelberg co-ordinate transformations
$\{\phi^{a}(x^{\mu})\}$ as
\begin{equation}\label{e21.42}
[\,f\,]^{\mu}_{\:\:\:\nu}\equiv
f_{\mu\nu}=\partial_{\mu}\phi^a\partial_{\nu}\phi^{b}\bar{f}_{ab}(\phi^{c}).
\end{equation}
Defining $\mathcal{G}^{\prime}=g\mathcal{G}^2$ with
$\mathcal{G}^{\prime}_{\mu\nu}\equiv
[\mathcal{G}^{\prime}]^{\mu}_{\:\:\:\nu}$ we can write
\eqref{e21.41} as
\begin{equation}\label{e21.43}
\partial_{\mu}\phi^a\partial_{\nu}\phi^{b}\bar{f}_{ab}(\phi^{c})=\mathcal{G}^{\prime}_{\mu\nu}.
\end{equation}
From this equation we deduce that $f$ must be a diagonal matrix as
also it was obvious from our construction. Now let us also assume
that the background metric $\bar{f}$ is also of the diagonal form.
To find solutions to these set of equations we will follow the
method developed in \cite{constmmgr,cosmommgr}. Firstly we choose
$\bar{f}$ as
\begin{equation}\label{e21.44}
\bar{f}=diag(f_{\texttt{00}},f_{ii}),\: \text{ for\quad} i=1,2,3.
\end{equation}
With this choice therein and bearing in mind that the right hand
side is a diagonal matrix, from \eqref{e21.43} we obtain
\begin{subequations}\label{e21.45}
\begin{align}
&\sum\limits_{a=\texttt{0}}^{\texttt{3}}\big(\partial_{\mu}\phi^a)^2f_{aa}(\phi^b)=\mathcal{G}^{\prime}_{aa},\quad\forall\mu,\notag\\
\tag{\ref{e21.45}}\\
&\sum\limits_{a=\texttt{0}}^{\texttt{3}}\partial_{\mu}\phi^a\partial_{\nu}\phi^{a}f_{aa}(\phi^b)=0,\quad\text{when}\quad\mu\neq\nu.\notag
\end{align}
\end{subequations}
Note that if we propose the condition
\begin{equation}\label{e21.46}
\partial_{\mu}\phi^a\partial_{\nu}\phi^{a}=0,\quad\forall
a,\:\:\:\text{and}\:\:\:\mu\neq\nu,
\end{equation}
then we can satisfy the second set of equations in \eqref{e21.45}.
Furthermore these equations namely \eqref{e21.46} can be solved by
demanding
\begin{equation}\label{e21.47}
\partial_{\mu\neq a}\phi^{a}=0,\quad\forall \mu,a,
\end{equation}
meaning that $\phi^{a}=\phi^{a}(x^a)$ only. On the other hand as
discussed in \cite{constmmgr,cosmommgr} when \eqref{e21.46} is
used in the first set of equations in \eqref{e21.45} one can
simplify the first set to the form
\begin{equation}\label{e21.48}
\big(\partial_{c}\phi^c)^2f_{cc}=\mathcal{G}^{\prime}_{cc},\quad\forall
c,\text{\quad no sum on $c$}.
\end{equation}
By taking square root on both sides these equations become
\begin{equation}\label{e21.5}
\partial_{c}\phi^{c}\sqrt{f_{cc}}=\pm\sqrt{\mathcal{G}^{\prime}_{cc}}.
\end{equation}
Finally if we choose the diagonal components of the background
metric $\bar{f}$ as
\begin{equation}\label{e21.6}
f_{cc}=\frac{\mathcal{G}^{\prime}_{cc}}{\big(F_c(x^c)\big)^2},
\end{equation}
so that
\begin{equation}\label{e21.7}
\bar{f}= \left(\begin{matrix}
\frac{\mathcal{G}^{\prime}_{00}}{(F_0(x^0))^2}&\texttt{0}&\texttt{0}&\texttt{0}
\\
\texttt{0}&\frac{\mathcal{G}^{\prime}_{11}}{(F_1(x^1))^2}&\texttt{0}&\texttt{0}\\\texttt{0}&\texttt{0}&\frac{\mathcal{G}^{\prime}_{22}}
{(F_2(x^2))^2}&\texttt{0}\\\texttt{0}&\texttt{0}
&\texttt{0}&\frac{\mathcal{G}^{\prime}_{33}}{(F_3(x^3))^2}\end{matrix}\right),
\end{equation}
then
\begin{equation}\label{e21.8}
\phi^{c}(x^{c}) =\pm \int F_{c}(x^{c})dx^{c},
\end{equation}
become the solutions of \eqref{e21.48}. Here we have introduced
the completely arbitrary integrable functions $F_a(x^a)$'s. These
St\"{u}ckelberg scalar field solutions when the background metric
is chosen as \eqref{e21.7} are the solutions of the ansatz
\eqref{e10}. We should also state that to be able to construct
\eqref{e21.7} explicitly one first has to specify the effective
energy-momentum tensor $\tilde{T}$ (which has to obey a
conservation equation as we will discuss below) then one has to
solve the diagonal metric from the metric sector. We see that the
form of our solutions justifies the diagonal matrix assumption we
have made for \eqref{e10} all through our analysis. On the other
hand for more general non-diagonal forms of $\sqrt{\Sigma}$, $g$,
and $\tilde{T}$ one would have a mixed-differential term-wise
coupling in the partial differential equations \eqref{e21.43}
which can not be so easily brought to a decoupled form. Those
equations may also admit solutions in principle however the reader
should appreciate that solving them would be more involved than
the simpler decoupling solution method we have discussed above.
Another essential loss of generality was choosing the background
metric diagonal for decoupling and consequently generating
solutions to these partial differential equations. Assuming
non-diagonal background metric forms would cause similar coupling
complications. We further note that all the conditions on $\Delta$
needed to obtain the real roots of \eqref{e19} contain the matrix
$D$ in them. For this reason in general the sign conditions on
$\Delta$ for the reality of the solutions (as $\Delta$ is a
functional of $\tilde{T}$ via $D$) depending on the particular
choice of $\tilde{T}$ may or may not bring restrictions on its
domain. Similarly as $\sqrt{\Sigma}$ is a functional of
$\tilde{T}$ too the valid domain of the solutions of the
background metric and the St\"{u}ckelberg fields may also be
restricted for certain class of solutions of \eqref{e19} and the
choice of $\tilde{T}$. However as we have mentioned before there
also exists the freedom of assigning $e_1=tr\sqrt{\Sigma}$ as a
compensation in the most general set-up to design solutions which
are free of these restrictions. Now on the other hand if we set
\begin{equation}\label{e22}
B^2-3AC=0,
\end{equation}
which puts no restrictions on the regions of validity of the
solutions and the effective source\footnote{ Eq. \eqref{e22} does
not result in the above mentioned restrictions on the domain of
validity of solutions as $\Delta_0$ does not depend on
$\tilde{T}$.} then it is guaranteed that there is one real matrix
solution to \eqref{e19} and it is
\begin{equation}\label{e23}
\sqrt{\Sigma}=-\frac{1}{3}A^{-1}\big(B+(\Delta_1)^{1/3}\big),
\end{equation}
where due to the condition \eqref{e22} we have
\begin{equation}\label{e24}
\Delta_1=-B^3+27A^2D.
\end{equation}
We remark that since all the matrices in the above relations are
diagonal the power operations can directly be applied on the
diagonal entries. Now unlike the more general solutions we have
discussed the condition \eqref{e22} will determine $e_1$. To see
this first note that from \eqref{e22} we get
\begin{equation}\label{e25}
e_2=\frac{1}{3}\big(e_1^2-\frac{\beta_2}{\beta_3}e_1+\frac{\beta_2^2}{\beta_3^2}-\frac{3\beta_1}{\beta_3}\big).
\end{equation}
Substituting \eqref{e18.2} into this equation we obtain
\begin{equation}\label{e26}
\beta_2e_1^2+\big(6\beta_1-\frac{\beta_2^2}{\beta_3}\big)e_1+\frac{\beta_2^3}{\beta_3^2}-\frac{3\beta_1\beta_2}{\beta_3}-9C_1+9\beta_0
+\frac{3}{2}C_2\tilde{T}^\mu_{\:\:\:\mu}-3C_2\tilde{\tau}^\mu_{\:\:\:\mu}=0,
\end{equation}
which is a quadratic equation for $e_1=tr\sqrt{\Sigma}$. The
discriminant of this equation is
\begin{equation}\label{e27}
\Delta^{\prime}=36C_1\beta_2+36\beta_1^2-\frac{3\beta_2^4}{\beta_3^2}-36\beta_0\beta_2-6C_2\beta_2\tilde{T}^\mu_{\:\:\:\mu}+
12C_2\beta_2\tilde{\tau}^\mu_{\:\:\:\mu}.
\end{equation}
In order to have a real root for \eqref{e26} we must have
\begin{equation}\label{e28}
\Delta^{\prime}\geq 0.
\end{equation}
This brings a constraint on the trace of the effective
energy-momentum tensor $\tilde{T}$ as well as the one for
$\tilde{\tau}$. We will see in the cosmological solution scheme of
the next section that choosing equality in \eqref{e28} will fix
the equation of state of the effective ideal fluid. In spite of
this restriction on the other hand this solution has physical
advantages as it will not cause additional constraints on the
building blocks of the cosmological solutions. However if equality
is not chosen close inspection shows that the free parameters in
\eqref{e27} can still be tuned to give solutions in physically
sensible domains. For example to enlarge the domain of validity of
the solutions when $C_1>0,\beta_2>0,C_2<0$ one can tune $\beta_2$
to small values and when $C_1<0,\beta_2<0,C_2>0$ one can tune the
norm of $\beta_2$ to high values to release the restrictions on
$\tilde{T}^\mu_{\:\:\:\mu}$. On the other hand choosing the
discriminant as zero will relax all the domain restrictions but
fix the form of $\tilde{T}$ in return. Now assuming \eqref{e28} is
satisfied, the real solutions to \eqref{e26} become
\begin{subequations}\label{e29}
\begin{align}
e_1=tr\sqrt{\Sigma}=&-\frac{3\beta_1}{\beta_2}+\frac{\beta_2}{2\beta_3}\pm\big[\frac{9C_1}{\beta_2}+9\big(\frac{\beta_1}{\beta_2}\big)^2-\frac{3}{4}
\big(\frac{\beta_2}{\beta_3}\big)^2\notag\\
&-9\frac{\beta_0}{\beta_2}-\frac{3C_2}{2\beta_2}\tilde{T}^\mu_{\:\:\:\mu}+\frac{3C_2}{\beta_2}\tilde{\tau}^\mu_{\:\:\:\mu}\big]^{1/2}\tag{\ref{e29}}.
\end{align}
\end{subequations}
This result together with \eqref{e18.2} fix the trace coefficients
in the solution \eqref{e23} of the ansatz equation \eqref{e19}.
Now we can explicitly compute $\sqrt{\Sigma}$. Again from
\eqref{e23} we have
\begin{equation}\label{e30}
\sqrt{\Sigma}=\mathcal{H},
\end{equation}
where we define the matrix
\begin{equation}\label{e31}
\mathcal{H}=\frac{1}{3}\big(\frac{\beta_2}{\beta_3}+e_1\big)\mathbf{1}_4-\frac{1}{3\beta_3}\big(\mathcal{F}\big)^{1/3},
\end{equation}
with
\begin{equation}\label{e32}
\mathcal{F}=\big(\beta_2+\beta_3e_1\big)^3\mathbf{1}_4+27\beta_3^2\big(C_2g^{-1}\tilde{T}+C_2g^{-1}\tilde{\tau}\big).
\end{equation}
Following the same track of solution route we have introduced
earlier in this section we find that the background metric
$\bar{f}$ becomes
\begin{equation}\label{e32.1}
\bar{f}=diag\big(\frac{\mathcal{H}^{\prime}_{00}}{\big(F_0(x^0)\big)^2},\frac{\mathcal{H}^{\prime}_{ii}}{\big(F_i(x^i)\big)^2}\big),
\end{equation}
where $\mathcal{H}^{\prime}=g\mathcal{H}^2$. We can also write
down the St\"{u}ckelberg scalar solutions again as
\begin{equation}\label{e32.2}
\phi^{a}(x^{a}) =\pm \int F_{a}(x^{a})dx^{a}.
\end{equation}
Although by now we can explicitly construct the scalar solutions
for the more general case in \eqref{e21.8} or the special one in
\eqref{e32.2} their being the solutions of the St\"{u}ckelberg
sector of \eqref{e1} is not guaranteed yet. To guarantee this we
must focus on the scalar field equations of \eqref{e1}. It can be
directly deduced from the metric equation \eqref{e4} that the
scalar field equations must be equivalent to the covariant
constancy condition
\begin{equation}\label{e45}
\nabla^\mu\mathcal{T}^S_{\mu\nu}=0,
\end{equation}
where $\nabla^\mu$ is the covariant derivative of the Levi-Civita
connection of $g$. If we substitute our ansatz \eqref{e10} in this
equation we get
\begin{equation}\label{e46}
\nabla^\mu\mathcal{T}^S_{\mu\nu}=\nabla^\mu\big[C_1m^2g_{\mu\nu}+C_2m^2\tilde{T}_{\mu\nu}+C_2m^2\tilde{\tau}_{\mu\nu}-\frac{1}{2}C_2m^2\tilde{T}^\rho_{\:\:\:\rho}
g_{\mu\nu}\big]=0.
\end{equation}
Then by using the metric compatibility $\nabla^\mu
g_{\alpha\beta}=0$ we obtain a constraint
\begin{equation}\label{e47}
\nabla^\mu\big[\tilde{T}_{\mu\nu}+\tilde{\tau}_{\mu\nu}-\frac{1}{2}\tilde{T}^\rho_{\:\:\:\rho}
g_{\mu\nu}\big]=0,
\end{equation}
which can be considered as a modified conservation or continuity
equation for the effective energy-momentum tensor $\tilde{T}$.
Thus finally we conclude that if one chooses $\tilde{T}$ in
\eqref{e10} as a solution of \eqref{e47} then for the background
metric \eqref{e21.7} or \eqref{e32.1} the St\"{u}ckelberg scalar
solutions \eqref{e21.8} or \eqref{e32.2} of \eqref{e10}
respectively become the scalar field solutions of the massive
gravity action \eqref{e1} together with the diagonal metric $g$ to
be solved from the metric sector which we will inspect for
cosmological cases next.
\section{FLRW Dynamics}
Now we turn our attention to the metric sector. If we substitute
the ansatz \eqref{e10} into the metric equation \eqref{e4} we get
the on-shell equation
\begin{equation}\label{e48}
R_{\mu\nu}-\frac{1}{2}Rg_{\mu\nu}-\tilde{\Lambda}
g_{\mu\nu}+C_2m^2\tilde{T}_{\mu\nu}+C_2m^2\tilde{\tau}_{\mu\nu}-\frac{1}{2}C_2m^2\tilde{T}^\rho_{\:\:\:\rho}
g_{\mu\nu}=G_NT^{M}_{\mu\nu},
\end{equation}
where we have defined the effective cosmological constant
\begin{equation}\label{e49}
\tilde{\Lambda}=\frac{1}{2}\Lambda-C_1m^2.
\end{equation}
We see that upon using the ansatz \eqref{e10} the metric sector is
completely decoupled from the scalars whose contribution is
truncated to the presence of an effective cosmological constant
and an energy-momentum tensor. Let us consider \eqref{e48} for the
cosmological FLRW metric
\begin{equation}\label{e50}
g=-dt^2+\frac{a^2(t)}{1-kr^2}dr^2+a^2(t)r^2 d\theta^2 +a^2(t)r^2
sin^2 \theta d\varphi^2,
\end{equation}
in the spatially spherical coordinates $\{t,r,\theta,\varphi\}$.
We should note at this point that the cosmological metric is
diagonal in this frame so that our analysis in the previous
section is applicable. For consistency with the isotropy and
homogeneity in \eqref{e50} we also choose the physical and the
effective sources in \eqref{e48} as ideal fluids so that
\begin{subequations}\label{e51}
\begin{align}
T^M_{\mu\nu}&=(\rho(t)+p(t))U_{\mu}U_{\nu}+p(t)g_{\mu\nu},\notag\\
\tag{\ref{e51}}\\
\tilde{T}_{\mu\nu}&=(\tilde{\rho}(t)+\tilde{p}(t))U_{\mu}U_{\nu}+\tilde{p}(t)g_{\mu\nu}.
\notag
\end{align}
\end{subequations}
From these definitions we can deduce the trace of $T^M$ and
$\tilde{T}$ as
\begin{subequations}\label{e52}
\begin{align}
&T^{M\:\mu}_{\mu}=g^{\mu\nu}T^M_{\mu\nu}=3p-\rho,\notag\\
\tag{\ref{e52}}\\
&\tilde{T}^\mu_{\:\:\:\mu}=g^{\mu\nu}\tilde{T}_{\mu\nu}=3\tilde{p}-\tilde{\rho}.
\notag
\end{align}
\end{subequations}
Also referring to its definition in \eqref{e11.1} via \eqref{e51}
$\tilde{\tau}$ can be computed as
\begin{equation}\label{e52.3}
\tilde{\tau}_{\mu\nu}=-\tilde{p}g_{\mu\nu},
\end{equation}
for which we have taken $\tilde{\rho},\tilde{p}$ to be linearly
independent with $g^{\mu\nu}$. Its trace becomes
\begin{equation}\label{e52.4}
\tilde{\tau}^\mu_{\:\:\:\mu}=-4\tilde{p}.
\end{equation}
If now we use the metric \eqref{e50} in \eqref{e48} a standard
computation which is slightly modified due to the extra terms in
\eqref{e48} gives the $tt$-component equation
\begin{equation}\label{e53}
\big(\frac{\dot{a}}{a}\big)^2+\frac{k}{a^2}=\frac{G_N}{3}\rho-\frac{C_2m^2}{6}\big(5\tilde{p}+\tilde{\rho}\big)-\frac{\tilde{\Lambda}}{3},
\end{equation}
which is the modified Friedmann equation. The three spatial
component $ii$-equations lead to the same form of equation
\begin{equation}\label{e54}
\frac{2\ddot{a}}{a}=-\big(\frac{\dot{a}}{a}\big)^2-\frac{k}{a^2}-G_Np-\frac{C_2m^2}{2}\big(3\tilde{p}
-\tilde{\rho}\big)-\tilde{\Lambda}.
\end{equation}
By using \eqref{e53} in \eqref{e54} we obtain the modified cosmic
acceleration equation
\begin{equation}\label{e55}
\frac{\ddot{a}}{a}=-\frac{G_N}{6}\big(3p+\rho\big)+\frac{C_2m^2}{3}\big(\tilde{\rho}-\tilde{p}\big)-\frac{\tilde{\Lambda}}{3}.
\end{equation}
For the FLRW metric \eqref{e50} the energy-momentum conservation
law
\begin{equation}\label{e56}
\nabla^\mu T^M_{\mu\nu}=0,
\end{equation}
of the physical matter leads to the continuity equation
\begin{equation}\label{e57}
\dot{\rho}=-\frac{3\dot{a}}{a}\big(p+\rho\big).
\end{equation}
On the other hand if we compute the modified conservation equation
\eqref{e47} for the effective energy-momentum tensor $\tilde{T}$
of the ideal fluid defined in \eqref{e51} by using the FLRW metric
\eqref{e50} then for the spatial components $\nu=i=\texttt{1,2,3}$
we get the usual null result. However the $t$-component of
\eqref{e47} yields
\begin{equation}\label{e57.1}
\frac{1}{2}\big(5\dot{\tilde{p}}+\dot{\tilde{\rho}}\big)=-\frac{3\dot{a}}{a}\big(\tilde{p}+\tilde{\rho}\big).
\end{equation}
Therefore like the one for the physical matter this modified
conservation equation couples our ansatz-originated effective
ideal fluid to the Friedmann equations. As the reader may realize
this continuity equation is different than the physical one
\eqref{e57}. The main reason for this is that we demand the
effective ideal fluid energy-momentum tensor to emerge from the
Lagrangian \eqref{e16} rather than the usual physical Lagrangian
of the perfect fluids which would be simply the effective pressure
and in which the first law of thermodynamics is implicitly casted.
For this reason the effective ideal fluids taking role in the
cosmological dynamics here can be called non-physical. Specifying
the equation of state of the physical and the effective matter
will enable us to solve for the scale factor $a(t)$, the physical
and the effective pressure and the energy densities. The solved
exact form of the effective energy-momentum tensor together with
the FLRW metric \eqref{e50} then can be used in the results of the
previous section to explicitly construct the background metric
$\bar{f}$ which generates these solutions. On the other hand, in
our solution scheme of the scalar sector presented in Section
three the St\"{u}ckelberg scalars and the effective source namely,
in general $\tilde{T}$ or in special for the cosmological case the
components $\{\tilde{\rho},\tilde{p}\}$ of it are independent from
each other. For the special class of diagonal solutions we have
constructed one first chooses the functions $\{F_{a}(x^{a})\}$' s,
then determines the St\"{u}ckelberg scalars from \eqref{e21.8} by
integrating these functions. After deciding the form of
$\tilde{T}$ (i.e introducing the ideal effective fluid arguments
$\{\tilde{\rho},\tilde{p}\}$ for the cosmological case) one later
solves the physical metric and
$\tilde{T}/\{\tilde{\rho},\tilde{p}\}$ from the metric, the
physical matter equations, and the conservation equation of
$\tilde{T}$ (which corresponds to the scalar field equations as we
have discussed in the previous section). Finally, one uses these
solutions namely, $g$, and $\tilde{T}/\{\tilde{\rho},\tilde{p}\}$
together with the functions $\{F_{a}(x^{a})\}$ to construct the
necessary background metric $\bar{f}$ in \eqref{e21.7} which
consistently leads to the derived solutions both in the
St\"{u}ckelberg and the metric sectors of the theory. Therefore we
see that in this branch of diagonal solutions for the cosmological
case the St\"{u}ckelberg scalars and the effective fluid
properties $\{\tilde{\rho},\tilde{p}\}$ are completely independent
of each other as the formers are arbitrary. On the other hand, in
the more general picture of solutions (at least for the ones again
diagonal in $\{\sqrt{\Sigma},g,\bar{f}\}$) one may follow a
different but a more challenging track to generate a broader class
of solutions. In this method for example for the cosmological
solutions one can first solve $\{\tilde{\rho},\tilde{p}\}$ (in
other general cases, the components of $\tilde{T}$) from the
Friedmann equations (modified Einstein equations), the physical
matter equations, and the conservation equation of $\tilde{T}$,
and now one can independently choose a background metric
$\bar{f}$, then one can solve the coupled partial differential
equations in \eqref{e21.43} to obtain the St\"{u}ckelberg scalars
in terms of $\{g,\bar{f},\tilde{\rho},\tilde{p}\}$. Hence, in this
more general solution scenario which we have not considered here
the St\"{u}ckelberg scalars become functions of the thermodynamic
state of the effective ideal fluid. We should also remark another
important point here. Although the equation of state of the
physical matter is subject to natural constraints we are
completely free to choose any form
$\tilde{p}=\textit{f}(\tilde{\rho})$ of it for the effective
matter case. Even non-physical effective ideal fluid choices are
possible provided they satisfy \eqref{e57.1} which is different
than the universal energy-momentum conservation law \eqref{e57} of
physical matter. However as we have discussed in the previous
section in spite of this large freedom we have bounds on the
effective energy-momentum tensor to have real solutions of the
reference metric and the scalar sector. On the other hand in the
special solution case if \eqref{e28} is saturated that is to say
if $\Delta^{\prime}=0$ then the solutions are valid for the entire
coordinate span but now we have to fix the equation of state of
the effective matter as
\begin{equation}\label{e58}
\tilde{p}=\frac{1}{11}\tilde{\rho}+C^{\prime},
\end{equation}
where
\begin{equation}\label{e59}
C^{\prime}=\frac{1}{11}\big[\frac{6C_1}{C_2}+\frac{6\beta_1^2}{C_2\beta_2}-\frac{\beta_2^3}{2C_2\beta_3^2}-\frac{6\beta_0}{C_2}\big].
\end{equation}
We have obtained \eqref{e58} by using \eqref{e52} and
\eqref{e52.4} in \eqref{e27} then by equating the result to zero.
Another way of obtaining real solutions in \eqref{e26} is to
equate the constant coefficient to zero. This leads to an equation
of state
\begin{equation}\label{e59.1}
\tilde{p}=\frac{1}{11}\tilde{\rho}+C^{\prime\prime},
\end{equation}
with
\begin{equation}\label{e59.2}
C^{\prime\prime}=\frac{2}{33C_2}\big[9C_1-9\beta_0-\frac{\beta_2^3}{\beta_3^2}+\frac{3\beta_1\beta_2}{\beta_3}\big],
\end{equation}
where we have used \eqref{e52} and \eqref{e52.4}. Now before we
conclude let us turn attention to the St\"{u}ckelberg sector
solutions of the special type which satisfy \eqref{e22} and which
accompany the cosmological metric solutions we have discussed
here. We have from \eqref{e51}
\begin{equation}\label{e60}
\tilde{T}= \left(\begin{matrix}
\tilde{\rho}&\texttt{0}&\texttt{0}&\texttt{0}
\\
\texttt{0}&\tilde{p}g_{\texttt{1}\texttt{1}}&\texttt{0}&\texttt{0}\\\texttt{0}&\texttt{0}&\tilde{p}g_{\texttt{2}\texttt{2}}
&\texttt{0}\\\texttt{0}&\texttt{0}
&\texttt{0}&\tilde{p}g_{\texttt{3}\texttt{3}}\end{matrix}\right),
\end{equation}
for the coordinate system defined in \eqref{e50}. So we get
\begin{equation}\label{e61}
g^{-1}\tilde{T}= \left(\begin{matrix}
-\tilde{\rho}&\texttt{0}&\texttt{0}&\texttt{0}
\\
\texttt{0}&\tilde{p}&\texttt{0}&\texttt{0}\\\texttt{0}&\texttt{0}&\tilde{p}
&\texttt{0}\\\texttt{0}&\texttt{0}
&\texttt{0}&\tilde{p}\end{matrix}\right).
\end{equation}
Also via \eqref{e52.3}
\begin{equation}\label{e62}
g^{-1}\tilde{\tau}= \left(\begin{matrix}
-\tilde{p}&\texttt{0}&\texttt{0}&\texttt{0}
\\
\texttt{0}&-\tilde{p}&\texttt{0}&\texttt{0}\\\texttt{0}&\texttt{0}&-\tilde{p}
&\texttt{0}\\\texttt{0}&\texttt{0}
&\texttt{0}&-\tilde{p}\end{matrix}\right).
\end{equation}
Therefore by inspecting \eqref{e31} and \eqref{e32} under these
identifications we find that
\begin{equation}\label{e63}
\mathcal{H^\prime}= \left(\begin{matrix}
\mathcal{H}^\prime_{00}&\texttt{0}&\texttt{0}&\texttt{0}
\\
\texttt{0}&\texttt{0}&\texttt{0}&\texttt{0}\\\texttt{0}&\texttt{0}&\texttt{0}
&\texttt{0}\\\texttt{0}&\texttt{0}
&\texttt{0}&\texttt{0}\end{matrix}\right),
\end{equation}
where
\begin{equation}\label{e64}
\mathcal{H}^\prime_{00}=-\frac{1}{9\beta_3^2}\bigg[\beta_2+\beta_3e_1+\big[-\big(\beta_2+\beta_3e_1\big)^3+27\beta_3^2C_2(\tilde{p}
+\tilde{\rho})\big]^{1/3}\bigg]^2.
\end{equation}
Thus from \eqref{e32.1} we conclude that the background metric
$\bar{f}$ is also of the form \eqref{e63} and it reads
\begin{equation}\label{e64.5}
\bar{f}=diag(\frac{\mathcal{H}^\prime_{00}}{(F_0(x^0))^2},0,0,0).
\end{equation}
Now if we analyze \eqref{e21.5} for our special solution
satisfying \eqref{e22} we observe that within the cosmological
solutions we have discussed in this section three of the
St\"{u}ckelberg scalars decouple from the metric sector and become
completely arbitrary with no role in the effective terms in the
metric equation. Thus in these special type of solutions the
construction of $\mathcal{T}^{S}$ in \eqref{e9} which enters into
the metric equation \eqref{e4} via the ansatz \eqref{e10} it
satisfies gets no contribution from $\phi^i$'s but only from
$\phi^0$ for generating the cosmological solutions of this
section. However this trivial solution set which is legitimate
analytically does not promise a physically acceptable case nor the
background metric \eqref{e64.5}. For this reason in this special
case for the St\"{u}ckelberg sector instead of choosing the
solutions of \eqref{e21.5} as in \eqref{e21.6} which determine the
background metric one should follow a different method in which
one should choose the components of the background metric
completely or partially and then solve for the scalars. For
example if one chooses
\begin{equation}\label{e64.6}
\bar{f}=diag(\frac{\mathcal{H}^\prime_{00}}{(F_0(x^0))^2},f_{11}(x^\mu),f_{22}(x^\mu),f_{33}(x^\mu)),
\end{equation}
where $\{f_{ii}(x^\mu)\}$ are arbitrary and
\begin{equation}\label{e64.7}
\phi^{0}(x^{0}) =\pm \int F_{0}(x^{0})dx^{0},\quad
\phi^{1},\phi^{2},\phi^{3}=\text{constant},
\end{equation}
then one can still satisfy \eqref{e21.5}. On the other hand beside
the above mentioned somewhat trivial cases we can find more
general solutions of $\{\phi^a\}$ for a completely specified
background metric by solving the equations in \eqref{e21.45} where
one should replace $\mathcal{G}^\prime$ by \eqref{e63}. Finally we
should remark at this point that for the more general case of
background metric and the St\"{u}ckelberg scalar solutions defined
via the equations \eqref{e21.1}-\eqref{e21.8} there appears no
problem of triviality occurring in the form of both $\bar{f}$ and
$\{\phi^a\}$'s like the one we have encountered above for the
solutions satisfying the condition \eqref{e22}.
\section{Conclusion}
We have constructed a specially chosen solution ansatz which has
enabled us to device a method to find solutions to the
St\"{u}ckelberg scalar sector of the massive gravity theory. We
have done this by decoupling the scalar sector from the metric one
and by reducing the task to finding the solutions of a mutual
constraint equation of the background metric and the scalars of
the theory. In this way we were able to find a broad solution
class of the St\"{u}ckelberg sector by specifying the associated
background metric in terms of the physical one and the ansatz
parametrizing effective matter energy-momentum tensor. In this
solution scheme the metric sector only sees the scalars as a
collective effect in the presence of an effective energy-momentum
tensor whose conservation equation is modified by a variation and
a trace term. Later we constructed the associated modified FLRW
cosmological dynamics by assigning an ideal fluid nature to both
the ordinary and the effective matter. We should emphasize the
point here that although the equations of different sectors are
decoupled from each other the solutions are intimately related.
One should first solve the physical metric and the effective
energy-momentum tensor from the metric and the modified
conservation equations then one should use these solutions to
construct the background metric which admits such a solution
scheme. Finally the solutions of the St\"{u}ckelberg scalars
trivially follow this construction up to a set of integrable
functions.

The cosmological solutions of the massive gravity is an active
topic of research which has reached some positive and negative
conclusions. For example it has been shown that for a flat
background metric there exist open FLRW solutions \cite{gum}
however there are no flat or closed FLRW solutions \cite{mc}. The
open FLRW solutions with the flat background metric have stability
problems \cite{st1,st2}. These facts led to the idea of choosing
different background metrics for the cosmological solutions such
as the de Sitter \cite{ln} and the FLRW type \cite{higuchi} ones.
However these solutions have non-physical consequences regarding
the Higuchi bound. On the other hand the inhomogeneous and
anisotropic cosmological solutions of the ghost-free massive
gravity have also been studied giving physically sensible results
in certain regions of the parameter space of the theory
\cite{ih1,ih2}. Contrary to the mainstream of the corresponding
literature we should remark that our approach does not
pre-determine the background metric but solves it for the
particular physical scenario in inspection. Therefore the
formalism of the present work addresses not only to a very rich
sector of the solution moduli of the background metric and the
scalars of the theory but also to a wide range of physical
applications via the choice of the effective matter. We certainly
owe this richness to the large solution space of the massive
gravity. We hope that the solution methodology achieved here will
provide an appropriate framework to generate new and physically
acceptable cosmological or astrophysical solutions of the theory.
However in this direction both the stability issues and the
Higuchi bound behavior of the solutions found here need to be
inspected in detail elsewhere.

We observe that the effective and collective contribution of the
scalars and the reference metric to the FLRW dynamics is
suppressed by a squared graviton mass coefficient. However there
also appear the $C_1$ and $C_2$ factors which possibly contain the
degree of freedom of tuning the solutions for agreement with the
physical needs. Such a scale generation may physically justify our
construction as an effective theory of sub-solutions of the the
exact theory. Both in the more general (which admits a wide range
of solutions depending on specifying the effective ideal fluid)
and the special cases we have mentioned, the modified cosmological
dynamics can separately be studied and their testable consequences
can further be classified as well. In this direction we should
remark that the solutions we have constructed do contain
self-accelerating ones. In particular, when we set $C_2$ to zero
then we have the solutions in which the contribution of the mass
sector to the metric one becomes solely an effective cosmological
constant (where still the richness of the corresponding
St\"{u}ckelberg scalar and the background metric solutions remains
intact) which is similar to many of such solutions constructed in
the literature. However the more interesting class of
self-accelerating solutions can be obtained by allowing the
existence of effective fluids. The reader may appreciate the
possibility of a wide range of such solutions as a consequence of
the free choice of the state equation for these fluids which in
general is completely arbitrary. For this reason we leave a
detailed examination of this issue for a later work. More
generally the point which deserves to be mentioned, and emphasized
on, is that, the present construction enables the freedom of
choosing any ordinary or exotic form of effective matter which
will admit a non-physical dynamics in generating solutions. This
is a consequence of the fact that although we have specified the
effective energy-momentum tensor in the ideal fluid form, while we
construct the on-shell Lagrangian corresponding to the solution
ansatz we did not take the usual Lagrangian of the ideal fluid
which would be just the effective pressure upon using the first
law of thermodynamics and which would generate the perfect fluid
energy-momentum tensor with no extra terms in the metric equation.
Thus our choice of the effective fluid does not obey the first law
of thermodynamics as can be obviously seen also from the modified
energy-momentum relation it satisfies and for this reason it can
be called non-physical. This brings the opportunity of generating
a rich class of solutions which would especially emerge from
effective fluid choices which can not have physical
correspondents. Furthermore the existence of the implicit solution
relation we have discussed above between the physical, the
background metrics and the effective matter also suggests a
coupling dynamics among them. Such a construction which could
include a dynamical nature for the background metric and a
reasonable origin for the effective matter can be searched within
the context of bi-metric gravity cosmological solutions
\cite{bm1,bm2,bm3,bm4}. One can also separately consider to extend
the ansatz we have used to generate similar solutions of
bigravity. Finally, we should point out the possibility of
modifying our ansatz to other forms which may lead to various
other solutions not necessarily cosmological within the same line
of reasoning.
\section*{Acknowledgements}
We thank Merete Lillemark for useful communications and comments.


\begin{thebibliography}{99}
\bibitem{fp}
  Fierz M., and Pauli W.
  \textit{``On relativistic wave equations for particles of arbitrary spin in an electromagnetic
  field}'',
1939   \textit{Proc. Roy. Soc. Lond.} A {\textbf173} 211.
\bibitem{BD1}
  Boulware D. G., and Deser S.
  \textit{``Can gravitation have a finite range?}'',
1972  \textit{Phys. Rev.} D {\textbf 6} 3368.
\bibitem{BD2}
  Boulware D. G., and Deser S.
  ``\textit{Inconsistency of finite range gravitation}'',
1972  \textit{Phys. Lett.} B {\textbf 40} 227.
\bibitem{dgrt1}
  de Rham C., and Gabadadze G.
  ``\textit{Generalization of the Fierz-Pauli Action}'',
   2010 \textit{Phys. Rev.} D {\textbf 82} 044020
  arXiv:1007.0443 [hep-th].
\bibitem{dgrt2}
  de Rham C., Gabadadze G., and Tolley A. J.
  ``\textit{Resummation of Massive Gravity}'',
  2011  \textit{Phys. Rev. Lett.}  {\textbf 106} 231101
  arXiv:1011.1232 [hep-th].
\bibitem{hr1}
  Hassan S. F., and Rosen R. A.
  ``\textit{On Non-Linear Actions for Massive Gravity}'',
  2011 \textit{JHEP} {\textbf 1107} 009
  arXiv:1103.6055 [hep-th].
\bibitem{hr2}
  Hassan S. F., and Rosen R. A.
  ``\textit{Resolving the Ghost Problem in non-Linear Massive
  Gravity}'',
2012   \textit{Phys. Rev. Lett.}  {\textbf 108} 041101
  arXiv:1106.3344 [hep-th].
\bibitem{hr3}
  Hassan S. F., Rosen R. A., and Schmidt-May A.
  ``\textit{Ghost-free Massive Gravity with a General Reference
  Metric}'',
 2012  \textit{JHEP} {\textbf 1202} 026
  arXiv:1109.3230 [hep-th].
\bibitem{derham}
de Rham C.
  \textit{``Massive Gravity}'',
  arXiv:1401.4173 [hep-th].
\bibitem{constmmgr}
Y$\i$lmaz N. T. \textit{``Constant curvature solutions of minimal
massive gravity''}, arXiv:1403.6014 [gr-qc].
\bibitem{cosmommgr}
Y$\i$lmaz N. T. \textit{``Effective Fluid FLRW Cosmologies of
Minimal Massive Gravity''}, arXiv:1404.3892 [hep-th].
\bibitem{bac}
Baccetti V., Martin-Moruno P., and Visser M.
  \textit{``Null Energy Condition violations in bimetric gravity''}, 2012
  \textit{JHEP} {\textbf 1208} 148
  arXiv:1206.3814 [gr-qc].
\bibitem{gum}
G\"{u}mr\"{u}k\c{c}\"{u}o\u{g}lu A. E., Lin C., and Mukohyama S.
\textit{``Open FRW universes and self-acceleration from nonlinear
massive gravity''}, 2011 \textit{JCAP} \textbf{1111} 030
arXiv:1109.3845 [hep-th].
\bibitem{mc}
D'Amico G., de Rham C., Dubovsky S., Gabadadze G., Pirtskhalava
D., et al. \textit{``Massive Cosmologies''}, 2011 \textit{Phys.
Rev.} \textbf{D84} 124046 arXiv:1108.5231 [hep-th].
\bibitem{st1}
De Felice A., G\"{u}mr\"{u}k\c{c}\"{u}o\u{g}lu A. E., and
Mukohyama S. \textit{``Massive gravity: nonlinear instability of
the homogeneous and isotropic
  universe''}, 2012
  \textit{Phys. Rev. Lett.} \textbf{109} 171101  arXiv:1206.2080 [hep-th].
\bibitem{st2}
Vakili B., and Khosravi N. \textit{``Classical and quantum massive
cosmology for the open FRW universe''}, 2012 \textit{Phys. Rev.}
\textbf{D85} 083529
  arXiv:1204.1456 [gr-qc].
\bibitem{ln}
  Langlois D., and Naruko A.
  \textit{``Cosmological solutions of massive gravity on de
  Sitter}'',
  2012 \textit{Class. Quant. Grav.} {\textbf 29} 202001 arXiv:1206.6810 [hep-th].
\bibitem{higuchi}
Fasiello M., and Tolley A. J. \textit{``Cosmological perturbations
in Massive Gravity and the Higuchi bound''}, 2012 \textit{JCAP}
\textbf{1211} 035 arXiv:1206.3852 [hep-th].
\bibitem{ih1}
Motohashi H., and Suyama T. \textit{``Self-accelerating Solutions
in Massive Gravity on Isotropic Reference Metric''}, 2012
\textit{Phys. Rev.} \textbf{D86} 081502 arXiv:1208.3019 [hep-th].
\bibitem{ih2}
Gratia P., Hu W., and Wyman M. \textit{``Self-accelerating Massive
Gravity: Bimetric Determinant Singularities''}, 2014 \textit{Phys.
Rev.} \textbf{D89}
  027502 arXiv:1309.5947 [hep-th].
\bibitem{bm1}
Volkov M. S. \textit{``Cosmological solutions with massive
gravitons in the bigravity theory''}, 2012 \textit{JHEP}
\textbf{1201} 035 arXiv:1110.6153 [hep-th].
\bibitem{bm2}
von Strauss M., Schmidt-May A., Enander J., Mortsell E., and
Hassan S. F. \textit{``Cosmological Solutions in Bimetric Gravity
and their Observational Tests''}, 2012 \textit{JCAP} \textbf{1203}
042 arXiv:1111.1655 [gr-qc].
\bibitem{bm3}
Volkov M. S., \textit{``Exact self-accelerating cosmologies in the
ghost-free bigravity and massive gravity''}, 2012 \textit{Phys.
Rev.} \textbf{D86} 061502 arXiv:1205.5713 [hep-th].
\bibitem{bm4}
Akrami Y., Koivisto T. S., and Sandstad M. \textit{``Accelerated
expansion from ghostfree bigravity: a statistical analysis with
improved generality''}, 2013 \textit{JHEP} \textbf{1303} 099
arXiv:1209.0457 [astro-ph.CO].
\end{thebibliography}
\end{document}